%% file: manuscript.tex
\documentclass[pdflatex,sn-mathphys-num]{sn-jnl}
\usepackage[T1]{fontenc}
\usepackage[utf8]{inputenc}
\usepackage{amsmath,amssymb,amsfonts}
\usepackage{graphicx}
\usepackage{booktabs,array}
\usepackage{xurl}
\hypersetup{hypertexnames=false,pdftitle={Forecasting Brand Mentions in LLM Answers: Frozen Predictions Evaluated from Hours to Weeks},pdfauthor={Dmitrij Żatuchin}}
\begin{document}

\title[Forecasting Brand Mentions in LLM Answers]{Forecasting Brand Mentions in LLM Answers: Frozen Predictions Evaluated from Hours to Weeks}

\author*[1]{\fnm{Dmitrij} \sur{\.Zatuchin}}\email{dmitrij@rankfor.ai}
\affil*[1]{\orgname{Rankfor.AI Sp. z o.o.}, \orgaddress{\city{Wroc\l{}aw}, \country{Poland}}}

\abstract{Companies and agencies now monitor which brands large language models (LLMs) name when buyers ask for recommendations. Answers vary between runs, so a single check can mislead and repeated checks cost money. We extend an earlier 20-day replication to ask how well past brand mentions predict later ones over seven weeks. We froze probabilistic forecasts before each of four prospective collection waves of Gemini 3.1 Pro answers to the same 54 buyer questions: nine question families in English, Estonian and Finnish, each for two markets, with four of 12 tracked brands per question. The waves were collected between 1.5 hours and 52 days after the history used by the earliest forecast, and each scored 648 brand and answer labels from 162 answers. A smoothed per-question history achieved a Brier score of 0.0599 at 1.5 hours, against 0.1131 for a per-brand rate and 0.25 for a constant 0.5. A pooled rate clipped to 0.05--0.95, named as the challenger before its first test, scored 0.0259 after 8 hours, 0.0269 after 30 days and 0.0328 after 52 days. Of the brand and question cells named in all or none of the earlier answers, 94.5\% to 97.5\% kept that pattern in each later wave, and none switched from always to never or back. With a search instruction, 100 of 162 answers returned search activity, compared with 1 of 540 earlier Pro answers; mentions stayed predictable (Brier 0.0563), and Pro history predicted a smaller model's answers less well (0.0787). Human coding agreed with 298/300 sampled automatic labels; two coders agreed on 99/100 shared items ($\kappa=0.98$). The results cover one provider family, a fixed candidate set and lexical mentions. They support per-question history as the baseline that any forecasting or monitoring method for LLM brand mentions should beat.}


\maketitle

\section{Introduction}\label{sec:intro}

Buyers increasingly ask AI assistants which supplier, software or shop to choose, and companies want to know whether their brand appears in those answers. The answers are generated anew each time, and the same question can return different lists. A marketing team that checks once may see a brand that the next answer omits, or miss one that usually appears. Checking often enough to see the pattern costs money, and every additional check raises the question of how much has really changed.

This paper treats the problem as forecasting. For a fixed question, language, market, model and brand, we estimate the probability that the brand is named in the next answer, using only answers collected earlier. We then test the estimates on answers collected after the forecasts were frozen, at intervals from 1.5 hours to 52 days. Predictability over weeks motivates testing less frequent or uncertainty-directed monitoring against a uniform schedule. Whether that strategy saves cost without missing important changes, including changes in previously settled cells, requires a separate prospective evaluation.

This expanded version builds on \emph{Twenty Days Later} \cite{zatuchin2026twentydays}, which reported the September waves and a follow-up collected on 13 October (D1). We retain its 20-day result and add follow-ups collected on 22 October and 13 November, search-instruction and smaller-model comparisons, and a larger human-validation exercise. The September observations and frozen forecasts are shared with the earlier version.

Per-question mention history predicted later mentions substantially better than a per-brand rate, including at 30 and 52 days. A pooled rate clipped to 0.05--0.95, chosen before its first test, scored best. That comparison jointly varies smoothing, history size and recency; their individual contributions remain unresolved. We also test an instruction to search and transfer to a smaller model from the same provider. An earlier exploratory experiment on unseen brands did not meet its validation criterion (Appendix~\ref{sec:v02}); the prospective results concern brands with observed answer history.

\section{Related work}\label{sec:related}

Repeated identical prompts can return different outputs even at fixed decoding settings. Ouyang et al.\ documented this for code generation \cite{ouyang2025nondeterminism}, and Atil et al.\ measured it across models and tasks at nominally deterministic settings \cite{atil2024nondeterminism}. Elazar et al.\ studied the consistency of factual predictions across paraphrases in pretrained language models \cite{elazar2021consistency}. Over longer periods, model behaviour can change with provider updates: Chen et al.\ found large changes between two dated versions of the same commercial model \cite{chen2023changing}. These studies motivate repeated measurement. Here we examine whether earlier answers forecast later ones for a fixed commercial question.

Work on generative engine optimisation asks how content changes affect the visibility of sources in generated answers \cite{aggarwal2023geo}. Our earlier studies measured brand inclusion and concentration in LLM recommendations across five industries and three systems \cite{zatuchin2026consideration}, proposed a protocol for repeated-query auditing of brand recommendations \cite{zatuchin2026dice}, audited brand and source discovery under repeated queries with finite samples \cite{zatuchin2026discovery}, separated the effects of query language and exit IP on commercial recommendations \cite{zatuchin2026market}, showed that automated sentiment scores of LLM responses are sensitive to the query language \cite{zatuchin2026querylanguage}, and documented concentration in repeated outputs for a task with no correct answer \cite{zatuchin2026lottery}. These studies motivate repeated measurement of brand mentions. The present analysis extends our earlier prospective forecasting study across longer intervals and additional collection conditions.

We score forecasts with the Brier score \cite{brier1950} and the logarithmic score, both strictly proper \cite{gneiting2007scoring}. Our uncertainty intervals resample nine question families; with so few clusters, bootstrap inference is approximate \cite{cameron2008bootstrap}.

\section{Study design}\label{sec:design}

\subsection{Questions, brands and target}

The questions cover three software categories: sales CRM, online shop platforms and project management tools. Each category has three question families, giving nine families. Every family was written in English, Estonian and Finnish and asked for a buyer in Estonia and for a buyer in Finland, giving 54 question contexts. Each context is evaluated against four candidate brands from its category, 12 brands in all: HubSpot, Zoho CRM, Microsoft Dynamics 365 and Insightly; WooCommerce, BigCommerce, MyCashflow and PrestaShop; ClickUp, Smartsheet, ProofHub and Wrike. The prompts name no brand and ask for up to five products with a short justification.

The target is a lexical mention. A brand counts as mentioned when its name or a listed alias appears in the answer text as a contiguous sequence of Unicode word tokens after NFKC normalisation and case folding. A mention includes recommendations, neutral listings, negative remarks, comparisons and names inside web addresses. Assessing recommendation quality, sentiment, factual accuracy and source support would require separate measures. Each answer yields four candidate labels, one per listed brand; labels from one answer, from translations of one question and from repeated runs are dependent.

\subsection{Collection}

The six evaluation waves in Table~\ref{tab:waves} used the Gemini API from Tallinn, Estonia. The main model was \texttt{gemini-3.1-pro-preview} with the Google Search tool enabled, low thinking effort, temperature 1.0, a maximum of 8,192 output tokens and no system instruction. The model chose whether to use the available search tool. Requests were not retried, and only complete, untruncated answers were scored. Table~\ref{tab:waves} lists the waves.

\input{tables/waves.tex}

The history for the first forecasts came from an earlier development wave (v0.2): 162 answers to the same 54 contexts, three per context. Wave v0.3 followed 1.5 hours after the last history answer. Wave v0.4 asked the same contexts of Pro and, interleaved in the same hours, of \texttt{gemini-3.8-flash}. Wave v0.5 asked Pro again with one added system instruction: ``Use Google Search to check current information before you answer.'' Waves v0.8 and v0.9 repeated the v0.3 conditions about 30 and 52 days after September.

\subsection{Prospective forecasts}

Forecasts for the six evaluation waves in Table~\ref{tab:waves} were fixed before collection, without external preregistration. The separate D1 replication reported an OSF preregistration (\url{https://osf.io/x8h3p/}) and its own decision rule \cite{zatuchin2026twentydays}. We retain D1 under that original analysis in Section~\ref{sec:d1}. The tables, figure and November human-validation sample use the six-wave panel; D1 was not used in the frozen histories for the later forecasts.

\subsection{Predictors}

All predictors give a probability that brand $b$ is named in an answer to question context $q$. Let $m$ and $n$ be mention and answer counts in the history.

\begin{itemize}
\item \emph{Constant}: 0.5.
\item \emph{Overall rate}: $g$, the share of positive labels in the history.
\item \emph{Brand history}: $p_b = (m_b + 20g)/(n_b + 20)$, pooling a brand's labels across its 18 contexts.
\item \emph{v0.3 question model}: $(m_{bq} + 5p_b)/(n_{bq} + 5)$ with the three v0.2 answers per cell, so a cell named in all three answers receives at most about 0.9.
\item \emph{Pooled rate with $k$ answers}: $\min(\max(m_{bq}/k, 0.05), 0.95)$ over the specified earlier Pro waves for the cell: $k = 6$ (v0.2 and v0.3, frozen for v0.4), $k = 9$ (adding v0.4, frozen for v0.5) and $k = 12$ (adding v0.8, frozen for v0.9).
\item \emph{Flash history}: the same pooled rate over three Flash answers from v0.4.
\item \emph{v0.2 context model}: a logistic model of question and candidate features fitted in the development experiment and never refitted.
\end{itemize}

The smoothing strengths 20 and 5 were fixed before v0.3, following an earlier retrospective two-wave analysis. The pooled rate entered as a challenger for v0.4 after a diagnostic on the v0.3 answers, described in Section~\ref{sec:short}; that diagnostic is development evidence, and v0.4 was the pooled rate's first test. The context model, question model and each fixed pooled predictor keep their frozen values in later waves, allowing comparisons of the same probabilities at different horizons. Overall-rate and brand-history baselines were updated before v0.4 and v0.5. Both later waves, v0.8 and v0.9, retained the v0.5 baseline values. The twelve-answer pooled predictor added v0.8 observations for its first evaluation in v0.9.

\subsection{Scoring and gates}

The Brier score is the mean squared difference between forecast and outcome; the log loss is the mean negative log probability of the outcome. Lower is better for both. The four Pro waves without the search instruction had prospective gates: the primary predictor had to score strictly lower than the constant, the overall rate and brand history on both metrics, with all 162 answers complete. The primary was the question-history model for v0.3 and v0.4, the nine-answer pooled rate for v0.8, and the twelve-answer pooled rate for v0.9. The Flash and search-instruction arms were descriptive transfer tests without pass/fail gates. Differences between predictors carry 95\% percentile intervals from 2,000 bootstrap resamples of the nine question families, keeping each family's translations, markets, repeats and candidates together and pairing the predictors on the same draws.

\section{Results}\label{sec:results}

\begin{figure}[t]
\centering
\includegraphics[width=\textwidth]{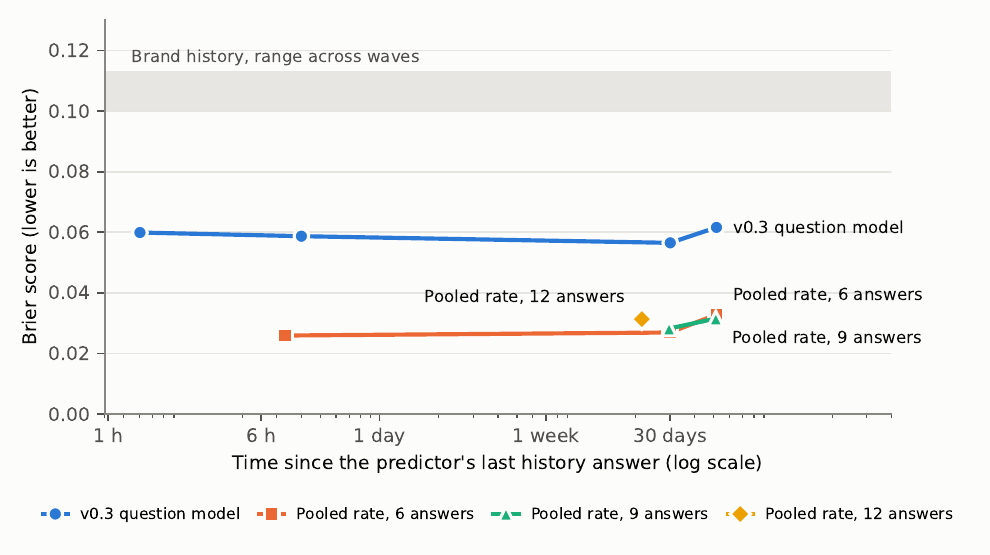}
\caption{Brier score of four frozen predictors against the time since the last history answer each predictor used. Lower is better; a constant 0.5 scores 0.25, above the plotted range. The grey band spans the brand-history scores of the four waves. Each point scores 648 candidate labels from 162 Gemini Pro answers.}\label{fig:horizon}
\end{figure}

Table~\ref{tab:pro} and Figure~\ref{fig:horizon} summarise all frozen predictors on the four Pro waves in the six-wave panel without the search instruction. Every wave passed its gate. The original D1 result is retained separately in Section~\ref{sec:d1}.

\input{tables/pro_scores.tex}

\subsection{Hours: history predicts the next answer}\label{sec:short}

At 1.5 hours (v0.3), the v0.3 question model achieved a Brier score of 0.0599 and a log loss of 0.2388, against 0.1131 and 0.3686 for brand history. The Brier difference was $-0.0532$, with paired interval $[-0.0858, -0.0249]$. The unchanged v0.2 context model came close at 0.0692, and its difference from the question model included zero ($[-0.0290, 0.0076]$).

A diagnostic on the v0.3 answers, run after scoring, showed where the question model lost accuracy. In the history, 200 of 216 brand and question cells were named in all three answers or in none, and none of those cells flipped between the two waves. About three quarters of the squared error came from such cells, where shrinkage toward the brand rate held forecasts away from 0 and 1. This led to the pooled rate as a challenger for v0.4. After 8 hours, the pooled rate over six answers scored 0.0259 and 0.1201, against 0.0587 and 0.2356 for the question model. Both paired intervals excluded zero in favour of the pooled rate (Brier $[0.0139, 0.0550]$; log loss $[0.0599, 0.1832]$ for the question model minus the pooled rate). The pooled rate also uses three more answers per cell, collected about 1.5 hours later, so this result combines lighter smoothing, more data and newer data.

\subsection{The original 20-day replication}\label{sec:d1}

The first version reported D1, collected on 13 October 2026, 20.3 days after v0.4 \cite{zatuchin2026twentydays}. Its 162 answers yielded 648 candidate labels. The nine-answer pooled forecast scored Brier 0.0281 and log loss 0.1264, against Brier 0.104 for brand history, 0.234 for the overall rate and 0.250 for a constant 0.5. It passed the original registered rule, including a Brier ceiling of approximately 0.0559 and the requirement that the returned model version match the requested version. It beat the three baselines in all nine question families (exact sign-flip $p=0.0039$).

Of 202 cells unanimous in v0.4, 194 remained unchanged at D1 and none reversed. The latest-wave-only comparator was close to the pooled rate (Brier 0.0292 versus 0.0281; paired family test $p=0.28$). No D1 answer returned a search query. These are the results reported in v1, retained with their original analysis; the additional waves below extend the time horizon.

\subsection{Weeks: the pattern held}\label{sec:weeks}

After about 30 days (v0.8), the six-answer pooled rate scored 0.0269 against its 0.0259 at 8 hours, and the primary nine-answer pooled rate scored 0.0282. The difference between the two pooled rates included zero ($[-0.0004, 0.0031]$), and the nine-answer rate beat the question model ($[-0.0488, -0.0117]$). After about 52 days (v0.9), the six-answer rate scored 0.0328 and the nine-answer rate 0.0315. The primary twelve-answer rate, which added the v0.8 answers, scored 0.0313; it matched the nine-answer rate within the uncertainty ($[-0.0035, 0.0020]$) and had a lower log loss than the six-answer rate (interval $[-0.0140, -0.0015]$). Scores remained low at both later follow-ups and below the brand-history band.

\input{tables/transitions.tex}

Table~\ref{tab:transitions} shows the same stability at the level of cells. Between the wave pairs shown, 94.5\% to 97.5\% of the cells that were named in all or none of the earlier answers kept that pattern, and no cell switched from always to never or back. Across the twelve Pro answers per cell in v0.2, v0.3, v0.4 and v0.8, 183 cells were named always or never; on 13 November, 178 of them were unchanged and none had flipped. The cells that moved concentrated in brands that were already named only some of the time. ClickUp, for example, appeared in 69\% of answers to the project management questions on 22 October and in 59\% on 13 November. Within each later wave, between 14 and 19 of 216 cells had three answers that disagreed.

\subsection{Search and model size}\label{sec:surface}

Without the instruction, Pro returned search activity in 1 of 540 answers across v0.2 to v0.4, in 2 of 162 in v0.8 and in 4 of 162 in v0.9; most responses therefore contained no returned search activity. With the search instruction, 100 of 162 answers returned search activity (Table~\ref{tab:surface}). Pro history still predicted mentions well: the nine-answer pooled rate scored 0.0563 and 0.2146, against 0.1079 and 0.3313 for brand history. Answers with returned search activity were harder to predict than those without it (Brier 0.0780 against 0.0211). This post hoc comparison is observational because the model chose when to search.

\input{tables/surface_scores.tex}

On 23 September 2026, Pro history predicted \texttt{gemini-3.8-flash} answers less well. The six-answer pooled rate scored 0.0787 on Flash answers, against 0.0259 on Pro answers collected in the same hours, and Flash returned search activity in 63 of 162 answers. History measured on one model is therefore a weaker guide to another model of the same family.

\subsection{Label validity}\label{sec:labels}

For each of the six panel waves, a deterministic sample of 24 candidate and answer pairs (12 automatic positives, 12 automatic negatives, distinct answers) was read twice by a second Gemini model in separate sessions. In every wave the two readings agreed with each other and with the automatic labels, except in v0.9, where one reading marked a brand absent because it searched only for a spaced alias; the answer named the brand three times, and the automatic label was correct. These model-assisted checks preceded the independent human validation. An automated substring audit of all 3,888 labels flagged three automatic negatives for possible inflected-name misses, all Estonian and all in the instructed wave. Human coding of these three cases is reported below.

The original D1 study reported a separate 96-label author check, with 94 initial agreements and two rater omissions identified with AI assistance and confirmed by the author \cite{zatuchin2026twentydays}. The independent two-coder exercise below covers the six-wave panel and uses a different sample.

A human validation protocol was fixed on 14 November 2026, before coding. The population comprised 3,888 candidate--answer pairs from six evaluated API waves, including 1,461 automatic positives. The main sample contained 300 pairs: 25 automatic positives and 25 negatives per wave, selected deterministically with language balancing and at most one pair per answer. The author coded all 300 pairs and the three targeted cases; a second team member independently coded 100 overlapping main-sample pairs in a different order. The author confirmed that both coders worked without AI assistance and were blinded to the automatic labels and each other's responses. They assessed brand presence, including inflected forms; recommendation quality and factual accuracy were outside the coding task.

On the main sample, the first coder confirmed 150/150 automatic positives (Wilson 95\% interval 97.50--100\%) and 148/150 automatic negatives (95.27--99.63\%); no main-sample item was marked unsure. The two discrepancies were in v0.4 Pro, one Estonian and one Finnish. Weighting agreement within each wave and automatic-label stratum by its population count gave descriptive estimates of 99.17\% accuracy, 100\% precision and 97.84\% recall relative to the first coder. These estimates assume the sampled pairs represent their strata; language balancing and the one-pair-per-answer constraint mean the wave/label weights do not capture every selection constraint. The targeted answers were excluded from the main sample but their strata retained full-population weights, so the estimates do not incorporate the separately identified targeted miss. The Wilson intervals apply to the sampled label groups; population uncertainty would require adjustment for the sampling design. Intercoder agreement was 99/100, with Cohen's $\kappa=0.98$; the remaining disagreement was not adjudicated.

Among the three purposively selected cases, the first coder marked one yes and two unsure. These cases were excluded from the general accuracy estimates and the two unsure cases remain unresolved. Some extraction errors and uncertainties remain. All forecast scores retain the original frozen automatic labels; no labels were changed by the human check.

\section{Discussion}\label{sec:discussion}

For a fixed question under these model and collection conditions, the brands an LLM names were highly predictable from the brands it named before, and the pattern held for seven weeks. Per-question history is therefore a strong baseline for forecasting AI brand mentions. A method that claims to forecast mentions, or a dashboard that reports a change, should show that it improves on what the same question returned before. The concentration of changes in mixed cells also suggests a monitoring strategy to test. Cells named always or never usually stayed that way between waves weeks apart, while changes concentrated in cells already named some of the time. This study did not compare monitoring allocations, measure cost savings or test whether sampling fewer settled cells would miss important changes.

The pooled predictors performed well at short and medium horizons. The post-hoc diagnostic is consistent with shrinkage toward the brand rate contributing to error on stable cells. The prospective pooled comparison also uses more and newer history, leaving the contribution of smoothing unresolved. Predictability also depended on the model and collection conditions: mentions remained predictable with the search instruction, while Pro history predicted Flash answers less well. Forecasts should therefore be evaluated under the conditions in which they will be used.

\subsection{Limitations}

The findings describe 12 known brands and 54 question contexts within one provider's API. Their applicability to consumer chat interfaces, logged-in sessions, other providers, unfamiliar brands and new questions remains to be tested. The preview model may also have changed during the study while retaining the same identifier.

The confidence intervals are approximate because they rely on nine question families. The follow-ups provide limited evidence about the timing or rate of change, so a longer observation period would be needed to establish how quickly forecasts lose accuracy.

The outcome is lexical brand presence, including incidental and negative mentions. Recommendation strength, factual accuracy and source support would require additional measures. Human validation found occasional extraction errors and two unresolved targeted cases (Section~\ref{sec:labels}). The reported forecast scores retain the automatic labels. Population-weighted validation estimates depend on how well the sampled pairs represent their strata; agreement on the 100 shared items provides evidence about that subset.

\section{Conclusion}\label{sec:conclusion}

Forecasts frozen from earlier answers predicted which of 12 brands Gemini 3.1 Pro would name for 54 buyer questions, at intervals from 1.5 hours to 52 days. A per-question pooled rate clipped to 0.05--0.95 scored best and lost little accuracy over seven weeks, and no settled brand and question cell switched sides. Per-question history should be the default comparator for forecasting and monitoring LLM brand mentions, and allocating more monitoring to unsettled cells is a hypothesis for future evaluation.

\section*{Declarations}

\textbf{Data availability.} Supporting data are available on Zenodo at \url{https://doi.org/10.5281/zenodo.24249785}. The deposit contains question texts, candidate aliases, frozen forecasts, automatic mention labels and aggregate evaluation and human-validation results for the six-wave panel. The original D1 data are archived at \url{https://doi.org/10.5281/zenodo.23351313}.

\textbf{Competing interests.} The author founded Rankfor.AI, which sells AI visibility measurement.

\textbf{Funding.} This work was funded by Rankfor.AI Sp. z o.o.

\textbf{Use of AI tools.} The author used Anthropic Claude and OpenAI Codex to support data collection and processing and manuscript preparation, including language and logical-structure checks. The author takes responsibility for the research, interpretation and final manuscript. Gemini models were evaluated in the study and used for the model-assisted label checks described above. The independent two-coder annotation was completed without AI assistance.

\bibliography{references}

\clearpage
\appendix
\section{Exploratory forecasting for unseen brands}\label{sec:v02}

An earlier development experiment (v0.2) tested whether official brand evidence could help forecast mentions for brands without prior answer history. It included 24 brands: 12 for training, 6 for validation and 6 reserved. Of five question families, three were used for training, one for validation and one was reserved. The 216 Pro answers yielded 648 training labels from 162 answers (247 positive) and 108 validation labels from 54 answers (85 positive).

Adding evidence features produced no improvement over the context model (Table~\ref{tab:v02}). All three predictors also scored worse than a constant 0.5 benchmark added after evaluation. Mention prevalence differed sharply between the training and validation sets: for project management, 37 of 216 training labels and all 36 validation labels were positive. These differences limit the interpretation of the transfer test. The experiment ended after this validation; the planned follow-up wave was cancelled.

\input{tables/v02.tex}

\end{document}

%% file: tables/waves.tex
\begin{table}[t]
\small
\caption{Six-wave evaluation panel and its initial history. Each wave asked the same 54 question contexts three times (162 answers, 648 candidate labels). Times are UTC, 2026. Search metadata: answers that returned at least one search query.}\label{tab:waves}
\setlength{\tabcolsep}{3pt}
\begin{tabular*}{\textwidth}{@{\extracolsep\fill}llllrr}
\toprule
Wave & Collected & Model & Condition & Positive labels & Searched \\
\midrule
v0.2 (history) & 22 Sep 19:57--20:15 & Pro & search allowed & 247 of 648 & 0 \\
v0.3 & 22 Sep 21:41--21:53 & Pro & search allowed & 246 of 648 & 1 \\
v0.4 Pro & 23 Sep 05:50--06:08 & Pro & search allowed & 244 of 648 & 0 \\
v0.4 Flash & same window & Flash & search allowed & 254 of 648 & 63 \\
v0.5 grounded & 23 Sep 08:38--08:55 & Pro & instructed to search & 223 of 648 & 100 \\
v0.8 & 22 Oct 20:56--21:08 & Pro & search allowed & 248 of 648 & 2 \\
v0.9 & 13 Nov 10:33--10:46 & Pro & search allowed & 246 of 648 & 4 \\
\botrule
\end{tabular*}
\footnotetext{Pro: \texttt{gemini-3.1-pro-preview}; Flash: \texttt{gemini-3.8-flash}. The separate 13 October D1 replication is described in Section~\ref{sec:d1}. The v0.2 history is the 162 answers of the training brands and question families of an earlier development wave of 216 calls.}
\end{table}

%% file: tables/pro_scores.tex
\begin{table}[t]
\footnotesize
\caption{Brier score / log loss of frozen predictors on later Gemini Pro answers (search allowed, no instruction). Lower is better. Each cell scores 648 candidate labels from 162 answers. A dash marks a predictor that did not exist yet, or had seen that wave.}\label{tab:pro}
\setlength{\tabcolsep}{3pt}
\begin{tabular*}{\textwidth}{@{\extracolsep\fill}lllll}
\toprule
Frozen predictor & v0.3 & v0.4 & v0.8 & v0.9 \\
\midrule
Constant 0.5 & 0.2500 / 0.6931 & 0.2500 / 0.6931 & 0.2500 / 0.6931 & 0.2500 / 0.6931 \\
Overall rate & 0.2355 / 0.6639 & 0.2348 / 0.6624 & 0.2363 / 0.6654 & 0.2355 / 0.6639 \\
Brand history & 0.1131 / 0.3686 & 0.1051 / 0.3335 & 0.0996 / 0.3099 & 0.1036 / 0.3204 \\
Flash history (3 answers) & -- & -- & 0.0763 / 0.2760 & 0.0735 / 0.2669 \\
v0.2 context model & 0.0692 / 0.2411 & 0.0723 / 0.2492 & 0.0637 / 0.2236 & 0.0711 / 0.2435 \\
v0.3 question model & 0.0599 / 0.2388 & 0.0587 / 0.2356 & 0.0565 / 0.2284 & 0.0616 / 0.2404 \\
Pooled rate, 6 answers & -- & 0.0259 / 0.1201 & 0.0269 / 0.1226 & 0.0328 / 0.1428 \\
Pooled rate, 9 answers & -- & -- & 0.0282 / 0.1264 & 0.0315 / 0.1384 \\
Pooled rate, 12 answers & -- & -- & -- & 0.0313 / 0.1361 \\
\midrule
Days since v0.2 history & 0.06 & 0.40 & 30.03 & 51.60 \\
\botrule
\end{tabular*}
\footnotetext{The overall rate and brand history were updated before v0.4 and v0.5, then carried unchanged into v0.8 and v0.9; the v0.2 context model, the v0.3 question model and each pooled rate keep the values frozen when they were created.}
\end{table}

%% file: tables/transitions.tex
\begin{table}[t]
\small
\caption{Stability of brand--question cells between the specified wave pairs. Settled: the brand was named in all three or none of the earlier wave's answers. Moved: a settled cell changed by one or two answers. Flipped: always became never, or the reverse. Mixed after: cells whose three answers in the later wave disagreed.}\label{tab:transitions}
\setlength{\tabcolsep}{3pt}
\begin{tabular*}{\textwidth}{@{\extracolsep\fill}lrrrrr}
\toprule
Interval & Settled before & Unchanged & Moved & Flipped & Mixed after \\
\midrule
v0.3 to v0.4 (8 hours) & 197 & 192 & 5 & 0 & 14 of 216 \\
v0.4 to v0.8 (29.6 days) & 202 & 193 & 9 & 0 & 15 of 216 \\
v0.8 to v0.9 (21.6 days) & 201 & 190 & 11 & 0 & 19 of 216 \\
\botrule
\end{tabular*}
\end{table}

%% file: tables/surface_scores.tex
\begin{table}[t]
\small
\caption{Brier score / log loss of Pro-history predictors on two other conditions, 23 September 2026. Lower is better; 648 candidate labels from 162 answers each.}\label{tab:surface}
\setlength{\tabcolsep}{3pt}
\begin{tabular*}{\textwidth}{@{\extracolsep\fill}lll}
\toprule
Frozen predictor & Flash answers & Pro instructed to search \\
\midrule
Constant 0.5 & 0.2500 / 0.6931 & 0.2500 / 0.6931 \\
Overall rate & 0.2385 / 0.6699 & 0.2269 / 0.6464 \\
Brand history & 0.1238 / 0.3820 & 0.1079 / 0.3313 \\
v0.2 context model & 0.0989 / 0.3258 & 0.0833 / 0.2718 \\
v0.3 question model & 0.0897 / 0.3096 & 0.0729 / 0.2655 \\
Pooled Pro rate & 0.0787 / 0.2837 & 0.0563 / 0.2146 \\
\botrule
\end{tabular*}
\footnotetext{The pooled Pro rate uses six answers per cell for the Flash wave and nine for the instructed wave, as frozen for each.}
\end{table}

%% file: tables/v02.tex
\begin{table}[t]
\small
\caption{Development experiment v0.2: forecasting unseen brands on a held-out question family. Validation: 108 candidate labels from 54 answers, 85 positive. Lower is better.}\label{tab:v02}
\setlength{\tabcolsep}{3pt}
\begin{tabular*}{\textwidth}{@{\extracolsep\fill}lrr}
\toprule
Predictor & Brier & Log loss \\
\midrule
Frequency baseline & 0.3323 & 0.8613 \\
Context model & 0.4885 & 1.3481 \\
Context model with evidence features & 0.4914 & 1.3570 \\
Constant 0.5 (post hoc) & 0.2500 & 0.6931 \\
\botrule
\end{tabular*}
\end{table}

%% file: manuscript.bbl

\begin{thebibliography}{15}
\ifx \bisbn   \undefined \def \bisbn  #1{ISBN #1}\fi
\ifx \binits  \undefined \def \binits#1{#1}\fi
\ifx \bauthor  \undefined \def \bauthor#1{#1}\fi
\ifx \batitle  \undefined \def \batitle#1{#1}\fi
\ifx \bjtitle  \undefined \def \bjtitle#1{#1}\fi
\ifx \bvolume  \undefined \def \bvolume#1{\textbf{#1}}\fi
\ifx \byear  \undefined \def \byear#1{#1}\fi
\ifx \bissue  \undefined \def \bissue#1{#1}\fi
\ifx \bfpage  \undefined \def \bfpage#1{#1}\fi
\ifx \blpage  \undefined \def \blpage #1{#1}\fi
\ifx \burl  \undefined \def \burl#1{\textsf{#1}}\fi
\ifx \doiurl  \undefined \def \doiurl#1{\url{https://doi.org/#1}}\fi
\ifx \betal  \undefined \def \betal{\textit{et al.}}\fi
\ifx \binstitute  \undefined \def \binstitute#1{#1}\fi
\ifx \binstitutionaled  \undefined \def \binstitutionaled#1{#1}\fi
\ifx \bctitle  \undefined \def \bctitle#1{#1}\fi
\ifx \beditor  \undefined \def \beditor#1{#1}\fi
\ifx \bpublisher  \undefined \def \bpublisher#1{#1}\fi
\ifx \bbtitle  \undefined \def \bbtitle#1{#1}\fi
\ifx \bedition  \undefined \def \bedition#1{#1}\fi
\ifx \bseriesno  \undefined \def \bseriesno#1{#1}\fi
\ifx \blocation  \undefined \def \blocation#1{#1}\fi
\ifx \bsertitle  \undefined \def \bsertitle#1{#1}\fi
\ifx \bsnm \undefined \def \bsnm#1{#1}\fi
\ifx \bsuffix \undefined \def \bsuffix#1{#1}\fi
\ifx \bparticle \undefined \def \bparticle#1{#1}\fi
\ifx \barticle \undefined \def \barticle#1{#1}\fi
\bibcommenthead
\ifx \bconfdate \undefined \def \bconfdate #1{#1}\fi
\ifx \botherref \undefined \def \botherref #1{#1}\fi
\ifx \url \undefined \def \url#1{\textsf{#1}}\fi
\ifx \bchapter \undefined \def \bchapter#1{#1}\fi
\ifx \bbook \undefined \def \bbook#1{#1}\fi
\ifx \bcomment \undefined \def \bcomment#1{#1}\fi
\ifx \oauthor \undefined \def \oauthor#1{#1}\fi
\ifx \citeauthoryear \undefined \def \citeauthoryear#1{#1}\fi
\ifx \endbibitem  \undefined \def \endbibitem {}\fi
\ifx \bconflocation  \undefined \def \bconflocation#1{#1}\fi
\ifx \arxivurl  \undefined \def \arxivurl#1{\textsf{#1}}\fi
\csname PreBibitemsHook\endcsname

\bibitem[\protect\citeauthoryear{{\.Z}atuchin}{2026}]{zatuchin2026twentydays}
\begin{botherref}
\oauthor{\bsnm{{\.Z}atuchin}, \binits{D.}}:
Twenty Days Later: A Pre-Registered Test of History-Based Forecasts of Brand
  Mentions in Gemini Answers.
Version 1, submitted 14 October 2026
(2026).
\url{https://arxiv.org/abs/2610.19608v1}
\end{botherref}
\endbibitem

\bibitem[\protect\citeauthoryear{Ouyang
  et~al.}{2025}]{ouyang2025nondeterminism}
\begin{barticle}
\bauthor{\bsnm{Ouyang}, \binits{S.}},
\bauthor{\bsnm{Zhang}, \binits{J.M.}},
\bauthor{\bsnm{Harman}, \binits{M.}},
\bauthor{\bsnm{Wang}, \binits{M.}}:
\batitle{An empirical study of the non-determinism of {ChatGPT} in code
  generation}.
\bjtitle{ACM Transactions on Software Engineering and Methodology}
\bvolume{34}(\bissue{2}),
\bfpage{1}--\blpage{28}
(\byear{2025})
\doiurl{10.1145/3697010}
\end{barticle}
\endbibitem

\bibitem[\protect\citeauthoryear{Atil et~al.}{2024}]{atil2024nondeterminism}
\begin{botherref}
\oauthor{\bsnm{Atil}, \binits{B.}},
\oauthor{\bsnm{Aykent}, \binits{S.}},
\oauthor{\bsnm{Chittams}, \binits{A.}},
\oauthor{\bsnm{Fu}, \binits{L.}},
\oauthor{\bsnm{Passonneau}, \binits{R.J.}},
\oauthor{\bsnm{Radcliffe}, \binits{E.}},
\oauthor{\bsnm{Rajagopal}, \binits{G.R.}},
\oauthor{\bsnm{Sloan}, \binits{A.}},
\oauthor{\bsnm{Tudrej}, \binits{T.}},
\oauthor{\bsnm{Ture}, \binits{F.}},
\oauthor{\bsnm{Wu}, \binits{Z.}},
\oauthor{\bsnm{Xu}, \binits{L.}},
\oauthor{\bsnm{Baldwin}, \binits{B.}}:
Non-Determinism of ``Deterministic'' {LLM} Settings.
arXiv:2408.04667
(2024).
\url{https://arxiv.org/abs/2408.04667}
\end{botherref}
\endbibitem

\bibitem[\protect\citeauthoryear{Elazar et~al.}{2021}]{elazar2021consistency}
\begin{barticle}
\bauthor{\bsnm{Elazar}, \binits{Y.}},
\bauthor{\bsnm{Kassner}, \binits{N.}},
\bauthor{\bsnm{Ravfogel}, \binits{S.}},
\bauthor{\bsnm{Ravichander}, \binits{A.}},
\bauthor{\bsnm{Hovy}, \binits{E.}},
\bauthor{\bsnm{Sch{\"u}tze}, \binits{H.}},
\bauthor{\bsnm{Goldberg}, \binits{Y.}}:
\batitle{Measuring and improving consistency in pretrained language models}.
\bjtitle{Transactions of the Association for Computational Linguistics}
\bvolume{9},
\bfpage{1012}--\blpage{1031}
(\byear{2021})
\doiurl{10.1162/tacl_a_00410}
\end{barticle}
\endbibitem

\bibitem[\protect\citeauthoryear{Chen et~al.}{2023}]{chen2023changing}
\begin{botherref}
\oauthor{\bsnm{Chen}, \binits{L.}},
\oauthor{\bsnm{Zaharia}, \binits{M.}},
\oauthor{\bsnm{Zou}, \binits{J.}}:
How is {ChatGPT}'s behavior changing over time?
arXiv:2307.09009
(2023).
\url{https://arxiv.org/abs/2307.09009}
\end{botherref}
\endbibitem

\bibitem[\protect\citeauthoryear{Aggarwal et~al.}{2023}]{aggarwal2023geo}
\begin{botherref}
\oauthor{\bsnm{Aggarwal}, \binits{P.}},
\oauthor{\bsnm{Murahari}, \binits{V.}},
\oauthor{\bsnm{Rajpurohit}, \binits{T.}},
\oauthor{\bsnm{Kalyan}, \binits{A.}},
\oauthor{\bsnm{Narasimhan}, \binits{K.}},
\oauthor{\bsnm{Deshpande}, \binits{A.}}:
{GEO}: Generative Engine Optimization.
arXiv:2311.09735
(2023).
\url{https://arxiv.org/abs/2311.09735}
\end{botherref}
\endbibitem

\bibitem[\protect\citeauthoryear{{\.Z}atuchin}{2026a}]{zatuchin2026consideration}
\begin{barticle}
\bauthor{\bsnm{{\.Z}atuchin}, \binits{D.}}:
\batitle{The synthesised consideration set: brand inclusion and concentration
  in large language model recommendations across five industries and three
  systems}.
\bjtitle{Journal of Marketing Analytics}
(\byear{2026})
\doiurl{10.1057/s41270-026-00562-3} .
\bcomment{Preprint: arXiv:2606.23057}
\end{barticle}
\endbibitem

\bibitem[\protect\citeauthoryear{{\.Z}atuchin}{2026b}]{zatuchin2026dice}
\begin{botherref}
\oauthor{\bsnm{{\.Z}atuchin}, \binits{D.}}:
The Dice Roll Method: A Standardized Protocol for Repeated-Query Auditing of
  Large Language Model Brand Recommendations.
arXiv:2609.04047
(2026).
\url{https://arxiv.org/abs/2609.04047}
\end{botherref}
\endbibitem

\bibitem[\protect\citeauthoryear{{\.Z}atuchin}{2026c}]{zatuchin2026discovery}
\begin{botherref}
\oauthor{\bsnm{{\.Z}atuchin}, \binits{D.}}:
Measuring Brand and Source Discovery under Repeated {LLM} Queries: A
  Finite-Sample Audit.
arXiv:2609.05059
(2026).
\url{https://arxiv.org/abs/2609.05059}
\end{botherref}
\endbibitem

\bibitem[\protect\citeauthoryear{{\.Z}atuchin}{2026d}]{zatuchin2026market}
\begin{botherref}
\oauthor{\bsnm{{\.Z}atuchin}, \binits{D.}}:
The Language of the Question Selects the Market: Query Language and Exit {IP}
  as Separable Factors in Commercial Recommendations from a Generative Search
  Interface.
arXiv:2608.30052
(2026).
\url{https://arxiv.org/abs/2608.30052}
\end{botherref}
\endbibitem

\bibitem[\protect\citeauthoryear{{\.Z}atuchin}{2026e}]{zatuchin2026querylanguage}
\begin{botherref}
\oauthor{\bsnm{{\.Z}atuchin}, \binits{D.}}:
Query Language Sensitivity in Automated Sentiment Scores of {LLM} Responses.
arXiv:2607.13304
(2026).
\url{https://arxiv.org/abs/2607.13304}
\end{botherref}
\endbibitem

\bibitem[\protect\citeauthoryear{{\.Z}atuchin}{2026f}]{zatuchin2026lottery}
\begin{botherref}
\oauthor{\bsnm{{\.Z}atuchin}, \binits{D.}}:
Ask a Language Model for Lottery Numbers: Concentration in Repeated Six-of-49
  Outputs.
arXiv:2610.00052
(2026).
\url{https://arxiv.org/abs/2610.00052}
\end{botherref}
\endbibitem

\bibitem[\protect\citeauthoryear{Brier}{1950}]{brier1950}
\begin{barticle}
\bauthor{\bsnm{Brier}, \binits{G.W.}}:
\batitle{Verification of forecasts expressed in terms of probability}.
\bjtitle{Monthly Weather Review}
\bvolume{78}(\bissue{1}),
\bfpage{1}--\blpage{3}
(\byear{1950})
\doiurl{10.1175/1520-0493(1950)078<0001:VOFEIT>2.0.CO;2}
\end{barticle}
\endbibitem

\bibitem[\protect\citeauthoryear{Gneiting and
  Raftery}{2007}]{gneiting2007scoring}
\begin{barticle}
\bauthor{\bsnm{Gneiting}, \binits{T.}},
\bauthor{\bsnm{Raftery}, \binits{A.E.}}:
\batitle{Strictly proper scoring rules, prediction, and estimation}.
\bjtitle{Journal of the American Statistical Association}
\bvolume{102}(\bissue{477}),
\bfpage{359}--\blpage{378}
(\byear{2007})
\doiurl{10.1198/016214506000001437}
\end{barticle}
\endbibitem

\bibitem[\protect\citeauthoryear{Cameron et~al.}{2008}]{cameron2008bootstrap}
\begin{barticle}
\bauthor{\bsnm{Cameron}, \binits{A.C.}},
\bauthor{\bsnm{Gelbach}, \binits{J.B.}},
\bauthor{\bsnm{Miller}, \binits{D.L.}}:
\batitle{Bootstrap-based improvements for inference with clustered errors}.
\bjtitle{Review of Economics and Statistics}
\bvolume{90}(\bissue{3}),
\bfpage{414}--\blpage{427}
(\byear{2008})
\doiurl{10.1162/rest.90.3.414}
\end{barticle}
\endbibitem

\end{thebibliography}
